\documentclass[copyright,creativecommons]{eptcs}

\providecommand{\event}{ICLP 2026}

\usepackage{packages}
\usepackage{macros}

\usepackage{amsmath}
\usepackage{graphicx}
\usepackage{multirow}
\usepackage{tabularray}
\usepackage{lineno}

\usepackage{iftex}

\usepackage[numbers]{natbib}

\ifpdf
  \usepackage{underscore}
  \usepackage[T1]{fontenc}
\else
  \usepackage{breakurl}
\fi

\title{Delayed Constraints in Narrowing for the Logic-Based Analyses of Real-Time Systems}

\author{
Santiago Escobar
\institute{
VRAIN, Universitat Polit\`ecnica de Val\`encia,  Spain
}
\email{sescobar@upv.es}
\and
Ra\'ul L\'opez-Rueda
\institute{
VRAIN, Universitat Polit\`ecnica de Val\`encia,  Spain
}
\email{rloprue@upv.es}
\and
Carlos Olarte
\institute{
LIPN, CNRS UMR 7030, Universit\'e Sorbonne Paris Nord, France
}
\email{olarte@lipn.univ-paris13.fr}
}

\def\titlerunning{Delayed Constraints in Narrowing for the Logic-Based Analyses of Real-Time Systems}
\def\authorrunning{S. Escobar, R. L\'opez-Rueda \& C. Olarte}

\begin{document}


\maketitle

\begin{abstract}
The formal analysis of real-time systems must address two dimensions of
infiniteness: an unbounded number of agents and messages, and a potentially
infinite state space induced by dense time. 
We present a novel narrowing-based
verification method that deals with both dimensions. Our approach integrates (i) rewriting modulo SMT for symbolic representation of
timing constraints, (ii) narrowing with logical variables to reason about
systems with an unknown number of agents, and (iii) a constraint
store over partially instantiated terms, in the style of constraint logic
programming. We further introduce a folding mechanism that, under certain
conditions, ensures termination of the symbolic analysis. The method has been
implemented as an extension of the Maude rewriting engine. We evaluate the approach by
verifying the correctness of a timed mutual exclusion protocol without
imposing bounds on the number of participating processes. Moreover, we show that
the framework uniformly supports the analysis of other real-time models,
including 
parametric timed automata with unspecified components that our method can synthesize. 
Our results suggest that the
proposed framework provides a sound and expressive basis for the
symbolic verification of real-time rewrite theories.

\end{abstract}

\renewcommand{\arraystretch}{1.5}

\section{Introduction}\label{sec:intro}
Rewriting logic 
(RL)~\cite{Mes92} 
is an expressive formalism
for the specification of a wide variety of systems, ranging from programming
languages and models of computation to protocols and distributed systems, among many others. The implementation of RL in the
rewriting engine \href{https://github.com/maude-lang/Maude}{Maude} has proven very practical for a number of verification tasks,
including invariant verification by checking the unreachability of a
pattern, representing the negation of the invariant,  in a  finite search space
obtained from a specific initial configuration. This is witnessed, for
instance, by RL-based systems  such as the
\href{https://kframework.org/}{K framework} (analysis of programming languages), \href{https://olveczky.se/RealTimeMaude/}{Real-Time Maude} (analysis of
real-time systems), the \href{https://carlosolarte.github.io/L-framework/}{L-framework} (analysis of deductive systems), 
and 
\href{https://personales.upv.es/sanesro/Maude-NPA_Protocols/index.html}{Maude-NPA}
(cryptographic protocol analysis), among others.

The advent of more powerful symbolic techniques, and their implementation in Maude (see \cite{DEEM+20}),
has opened the possibility of dealing with more challenging verification tasks,
including invariant verification but now for arbitrary SMT variables, appearing in the transition rules as fresh variables and both in the initial configuration and the invariant.
In fact, although real-time systems can be naturally specified as (real-time)
rewrite theories, the only analysis techniques available until the end of the last decade
were based on \emph{sampling} the system behavior.
While this approach has allowed the verification of complex real-time systems~\citep{Bae2026},
such analyses are in general neither sound nor complete when dense time is involved.
By combining rewriting of terms containing Boolean SMT expressions, 
and allowing rewrite rules to be guarded by such expressions,
it has recently become possible to provide sound and complete analysis methods in RL
for real-time formalisms such as parametric timed automata (PTA) and time Petri nets (\cite{ftscs-journal,DBLP:journals/fuin/AriasBOOP24}). 
In some cases, the resulting Maude implementations outperformed state-of-the-art tools
for these formalisms, while keeping a very natural and declarative specification style
based on rewrite rules.
However, despite this significant advance in the quest for better verification tools,
the theory and practice of RL/Maude constrained the use of these techniques
to systems in which all components are fully specified, that is,
to cases where the initial configuration is a \emph{ground} term.

Another major advance of the last and present decade was to endow Maude 
with 
logical variables
(\cite{DEEM+20}).
On the one hand, unification is a key mechanism in constraint logic programming,  
and equational unification
has steadily increased its use in new different areas
thanks to its greater efficiency.
The latest Maude 3.5.1 supports: 
(i) unification modulo the associativity and/or commutativity and/or identity axioms,
and
(ii) narrowing-based unification modulo \emph{user-definable} oriented equations and the previous axioms but having the \emph{finite variant property}.
On the other hand, concurrent systems are specified in Maude
using transition rules modulo an equational theory
and, when such transition rules are \emph{topmost}, 
narrowing provides a symbolic reachability
analysis method for \emph{infinite-state systems},
\ie{} 
\emph{can an instance of state $u$ reach an instance of state $v$ modulo the equational theory?} 
We can perform invariant verification but now for  arbitrary logical variables (as in Logic Programming), appearing in the transition rules as fresh variables and both in the initial configuration and the invariant.
However, 
this provides a semi-decision procedure when the narrowing-based search space is infinite.

Maude 3.5.1 also supports a powerful symbolic state space reduction that removes a symbolic  state $v'$ 
if it is an instance of a previously explored state modulo the equational theory.
This \emph{folding} technique \citep{DBLP:conf/rta/EscobarM07}
is based on the unfold/fold program manipulation and transformation approach of
\cite{Burstall-Darlington-77}
but adapted 
from functional programming
to model checking
using logic programming concepts.
It has proved useful in obtaining a finite narrowing-based search space in many situations, 
thus providing a decision procedure for invariant verification of infinite-state systems.
An extension of such folding narrowing to
rewrite rules guarded by Boolean SMT expressions is also available,
although not natively in Maude.

This paper leverages the above-mentioned techniques so that
invariant verification is performed now for  arbitrary (logical and SMT) variables, appearing in the transition rules as fresh variables and both in the initial configuration and the invariant, while still producing a finite search space.
For that, we take inspiration from a well-known approach from (constraint) logic programming, namely,
the use of \emph{delayed} (or suspended) constraints, which are accumulated and only solved
or propagated once they are further instantiated.
Our first contribution is the definition of \emph{delayed folding narrowing} where:
(i) conditions in rewrite rules may include SMT expressions with delayed parts;
(ii) right-hand sides in rewrite rules (RHS) may include variables not occurring in the left-hand side (LHS);
(iii) conditions may include variables not from the LHS and the RHS;
(iv) queries may contain (shared) variables in the initial and target states;
and
(v) queries may contain an initial SMT expression with delayed parts.

Our  framework enables
verification tasks that go beyond the analyses already possible with ``standard'' 
Maude. 
Our second contribution is to automatically verify, for the first time 
\footnote{Mutual exclusion 
was
verified for an arbitrary number of processes 
in \cite{OgataSICE2020,GhilardiNFM2012}.
In \cite{OgataSICE2020}, 
auxiliary lemmas were necessary
and no arbitrary
parameters $\gamma$ and $\delta$ were considered, as we do
in~\Cref{sec:fischer}. 
In \cite{GhilardiNFM2012}, 
arbitrary parameters were considered but
auxiliary lemmas were necessary and some form of acceleration to compress an unbounded number of transitions into one.
Other approaches 
using
UPPAAL or similar timed-automata model checkers
assume a fixed number of
processes and fixed time bounds. In contrast, 
IMITATOR  (\cite{Andre21}) allows parametric time parameters, but only
for a fixed number of processes.} , the correctness of the timed Fischer
mutual exclusion
protocol in its most general setting, 
for an arbitrary number of processes and arbitrary timed  parameters.
Key aspects of this verification task are
the 
design of a hierarchy of sorts to guarantee termination of the folding
procedure,  
 the use of logical variables to represent an
unspecified number of processes, and the  role played by rewrite rules
with delayed constraints.

As the 
third contribution,
we consider the timed dining philosophers system and show that,
by using logical variables, we can leave the controller of the system unspecified.
Our narrowing procedure can \emph{synthesize} such a controller,
that is, determining the missing transitions of the controller 
so that a given reachability property is satisfied.
For these two applications, we specify a general (extended)
real-time rewrite theory. As a result, our methods are applicable to a wide
range of other protocols and systems.

\noindent\emph{Organization.}  After the necessary preliminaries in \Cref{sec:prelim}, our 
folding narrowing procedure with delayed
constraints is explained in \Cref{sec:narrowing}. \Cref{sec:theory} introduces
what we call \emph{logical} real-time rewrite theories, which, unlike
``standard'' real-time rewrite theories, allow extended and delayed SMT
constraints in rule guards. This section also presents 
the two case studies. Related work is discussed
in \Cref{sec:related}, and we conclude in \Cref{sec:conc}. The
\href{https://depot.lipn.univ-paris13.fr/real-time-maude/logical-rt-maude}{companion repository} (\cite{tool}) contains the implementation of the system, the
case studies, and additional examples omitted due to
space limitations.

\section{Preliminaries}\label{sec:prelim}

This section gives the necessary background on rewriting logic (\cite{Mes92})
and its implementation in the Maude rewriting engine (\cite{maude-manual}).

A \emph{rewrite theory} (\cite{Mes92}) is 
a tuple $\mathcal{R} = (\Sigma, E, L, R)$
where: $\Sigma$ is an order-sorted signature that declares sorts, subsorts, and function symbols;
$E$ is a set of (conditional) 
    equalities of the form $t=t' \mbox{ \textbf{if} } \psi$,
    where 
    $t$ and $t'$ are terms of the same sort,
    and $\psi$ is a conjunction of equalities;
    $L$ is a set of \emph{labels};
    and
    $R$ is a set of labeled (conditional) 
    rewrite rules
    of the form
    $l : q \longrightarrow r \mbox{ \textbf{if} } \psi$,
    where $l \in L$ is a label,
    $q$ and $r$ are terms of the same sort,
    and
    $\psi$ is a conjunction of equalities. 

$T_{\Sigma, s}$ denotes the set of ground (\ie{} not containing variables)
terms of sort $s$,
 and $T_{\Sigma}(X)_s$  the set of 
 terms of sort $s$
 over a set of  sorted 
 variables $X$. $T_{\Sigma}(X)$ and
 $T_{\Sigma}$ denote all terms and ground terms, respectively.
The set of variables of a term $t$ is denoted by $\ovars{t}$.
 If $\sigma : X \rightarrow T_{\Sigma}(X)$
 is a substitution (or a \emph{ground} substitution $\sigma : X \rightarrow T_{\Sigma}$),  
  then $t \sigma$ 
  denotes 
 the term obtained
by simultaneously replacing each variable $x$ in $t$ with $\sigma(x)$.
The domain and range of a substitution are defined as expected. 
The restriction of the domain of a substitution $\sigma$ to a set of variables $W$ is represented as $\sigma|_{W}$.
%

An \textit{equation} is an unoriented pair $t = t'$, where $t,t' \in T_{\Sigma}(X)_s$ for some sort $\sort{s}\in\sort{S}$.   Given
$T_{\Sigma}(X)$ and a set $E$ of $T_{\Sigma}(X)$-equations, order-sorted equational logic induces a congruence relation $=_E$ on terms $t,t' \in T_{\Sigma}(X)$. 
%
%
An \emph{equational theory} $(\Sigma,E)$ is a pair with $\Sigma$ an order-sorted signature and $E$ a set of $T_{\Sigma}(X)$-equations. 
An \textit{$E$-unifier} for a $\Sigma$-equation $t = t'$ is a substitution $\sigma$ such that $t\sigma =_E t'\sigma$.  For $\ovars{t}\cup\ovars{t'} \subseteq W$, a set of substitutions $\textit{CSU}(t = t')_E^W$ is said to be a \textit{complete} set of unifiers for the 
$\Sigma$-equation $t = t'$ modulo $E$ away from $W$ iff: (i) each $\sigma \in \textit{CSU}(t = t')_E^W$ is an $E$-unifier of $t = t'$; (ii) for any $E$-unifier $\rho$ of $t = t'$ there is a substitution $\sigma \in \textit{CSU}(t = t')_E^W$ such that $\sigma|_W \sqsupseteq_{E} \rho|_W$ (i.e., there is a substitution $\eta$ such that $(\sigma\eta)|_{W} =_E \rho|_{W}$); and (iii) for all $\sigma \in \textit{CSU}(t = t')_E^W$, $\mathit{dom}(\sigma) \subseteq (\ovars{t}\cup\ovars{t'})$ and $\mathit{range}(\sigma) \cap W = \emptyset$.

%

A \emph{one-step rewrite}   $t \longrightarrow_{R,E} t'$,  holds 
if there are
a rule $l : q \longrightarrow r \mbox{ \textbf{if} } \psi$,
a subterm $u$ of $t$,
and a substitution $\sigma$ s.t. 
$u =_E q\sigma$,
$t'$ is the term obtained from $t$
by replacing 
$u$ with $r\sigma$,
and $v\sigma = v'\sigma$ holds
for each 
$v = v'$ in $\psi$.
We denote by
$\longrightarrow_{R,E}^\ast$
the reflexive-transitive closure of $\longrightarrow_{R,E}$.
A rewrite theory $\mathcal{R}$ is called \emph{topmost} iff there is a sort
$\mathit{State}$ at the top of one of the connected components of the subsort
partial order such that for each rule $l : q \longrightarrow r \mbox{
\textbf{if} } \psi$, both $q$ and $r$ have the top sort $\mathit{State}$, and
no operator has sort $\mathit{State}$ or any of its subsorts as an argument
sort. This intuitively means that (nondeterministic) computation specified by the
rules happens at the top of the state terms. 
We note that most distributed systems can be specified in rewriting logic in this way.

A Maude functional module (\texttt{{fmod}} \textit{M} \texttt{{is}} ... \texttt{{endfm}}) specifies an equational theory, and a Maude system module (\texttt{{mod}} \textit{M} \texttt{{is}} ... \texttt{{endm}}) specifies a rewrite theory.
Sorts and subsort relations are declared by the keywords
\texttt{{sort}}  and \texttt{{subsort}}.
For each sort $s$, Maude extends the subsort partial order with a super sort \lstinline[mathescape]{[$s$]} called \emph{kind}, 
and it represents undefined or error terms of sort $s$.
Function symbols, or \emph{operators}, are introduced as \texttt{{op}} $f$
\texttt{:} $s_1$ ... $s_n$ $\to$ $s$, where $s_1$, \ldots,  $s_n$ are the sorts
of its arguments, and $s$ is the sort of the returned value. Operators can have
user-definable syntax, with underbars `\verb@_@' marking each of the argument
positions (\eg{}  \verb@_+_@). Some operators can have equational attributes,
such as \texttt{{assoc}}, \texttt{{comm}}, and
\texttt{{id:}}$\;\iota$, stating that the operator is,
respectively, associative,  commutative,  and/or has identity element $\iota$.
An operator can also be declared to be a constructor (keyword \lstinline{ctor}) that
defines the data elements of its sort.
(Conditional) Equations and (conditional) rewrite rules are specified, respectively,  with the syntax
\texttt{{eq}} $t$ \texttt{=} $t'$, 
\texttt{{ceq}} $t$ \texttt{=} $t'$
\texttt{{if}} $\psi$, \lstinline[mathescape]|rl [$\mathit{label}$] $t$ => $t'$| and
\lstinline[mathescape]|crl [$\mathit{label}$] $t$ => $t'$ if $\phi$|.

For a signature $\Sigma$ and 
a set of equations $E$,
a \emph{built-in theory} $\mathcal{E}_0$
is a first-order theory with a signature $\Sigma_0 \subseteq \Sigma$,
where
(1) each sort $s$ in $\Sigma_0$ is minimal in $\Sigma$;
(2) $s \notin \Sigma_0$ for each operator $f:s_1\times \cdots \times s_n \rightarrow s$
    in $\Sigma \setminus \Sigma_0$; and
(3) $f$ has no other subsort-overloaded typing in $\Sigma_0$.
The satisfiability of a constraint in $\mathcal{E}_0$
is assumed to be decidable
using the SMT theory $\mathcal{T}_{\mathcal{E}_0}$
and we assume it is
consistent with  $(\Sigma, E)$, \ie{} 
for $\Sigma_0$-terms $t_1$ and $t_2$,
$\mathcal{T}_{\mathcal{E}} \models t_1 = t_2 \Leftrightarrow \mathcal{T}_{\mathcal{E}_0} \models t_1 = t_2$.

A \emph{constrained term} is a pair $\phi \parallel t$ of an SMT expression
$\phi$ in $\mathcal{E}_0$ and a term $t \in T_{\Sigma}(X_0)$ over variables
$X_0 \subseteq X$ of the built-in sorts in $\mathcal{E}_0$.
It symbolically represents all the ground instances of $t$ satisfying $\phi$: $\llbracket \phi \parallel t \rrbracket = \{t' \mid t' =_E t\sigma \ \mbox{and}\  \mathcal{T}_{\mathcal{E}_0} \models \phi\sigma \
\mbox{for }\ \sigma : X_0 \to T_{\Sigma_0} \}. $

For 
$t \in T_{\Sigma}(X)$,
an \emph{abstraction of built-ins}
is a pair $(t^\circ, \sigma^\circ)$
of 
a term $t^\circ \in T_{\Sigma \setminus \Sigma_0}(X)$
and
a substitution $\sigma^\circ : X_0 \to T_{\Sigma_0}(X_0)$
such that
$X_0 \subseteq X$,
$t = t^\circ \sigma^\circ$
and  $t^\circ$  contains no duplicate variables in $X_0$.
That is, any non-variable built-in subterms of $t$ are 
replaced by distinct built-in variables in $t^\circ$
and
new equalities are generated
$\Psi_{\sigma^\circ} = \bigwedge_{x \in \mathit{dom}(\sigma^\circ)} (x = x \sigma^\circ$).
Let $\phi \parallel t$ be a constrained term 
and $(t^\circ, \sigma^\circ)$ an
abstraction of built-ins for $t$.
If $\mathit{dom}(\sigma^\circ) \cap \ovars{\phi \parallel t} = \emptyset$,
then
$\llbracket \phi \parallel t \rrbracket
= 
\llbracket \phi \wedge \Psi_{\sigma^\circ} \parallel t^\circ \rrbracket$ 
(\cite{rocha-rewsmtjlamp-2017}).

Let $\mathcal{R}=(\Sigma, E, L, R)$ be a topmost theory
with a built-in theory $(\Sigma_0,E_0)$
s.t. for each rule $l : q \longrightarrow r \mbox{ \textbf{if} }
\psi$ in $R$, extra variables not occurring in the LHS $q$ 
are in $X_0$,
and
$\psi$ is a set of $\Sigma_0$ equalities. For instance,  as in the conditional rule 
\lstinline|crl p(x) + clk(u) => p(x) + m(t) + clk(u) if t > u|,
where \texttt{+} is associative and commutative, 
representing an agent
\lstinline{p(x)}  that 
creates a message
\lstinline{m(t)} where the new SMT variable  
\lstinline{t} is constrained by the global clock \lstinline{u}.
A \emph{one-step symbolic rewrite} $\phi \parallel t
\rightsquigarrow_{\mathcal{R},E} \phi' \parallel t'$ holds
iff there exist
a rule $l : q \longrightarrow r \mbox{ \textbf{if} } \psi$
and
a substitution $\sigma : X \to T_{\Sigma}(X_0)$
such that
(1)~$t =_E q\sigma$
and
$t' = r\sigma$,
(2)~$\mathcal{T}_{\mathcal{E}_0} \models (\phi \wedge  \psi
      \sigma) \Leftrightarrow \phi'$, and
(3)~$\phi'$ is $\mathcal{T}_{\mathcal{E}_0}$-satisfiable.

A \emph{symbolic rewrite}
on constrained terms
symbolically represents 
a (possibly infinite) set of system transitions.
Symbolic and concrete transitions are in tight correspondence: 
If $\phi_t \parallel t \rightsquigarrow^\ast \phi_u \parallel u$
is a symbolic rewrite,
then there exists a ``concrete'' rewrite  $t' \longrightarrow^\ast u'$
 with $t' \in \llbracket \phi_t \parallel t \rrbracket$
 and $u' \in \llbracket \phi_u \parallel u \rrbracket$.
Conversely,
for any concrete
rewrite $t' \longrightarrow^\ast u'$ with
$t' \in \llbracket \phi_t \parallel t \rrbracket$,
there exists
a symbolic rewrite $\phi_t \parallel t \rightsquigarrow^\ast \phi_u \parallel u$
with $u' \in \llbracket \phi_u \parallel u \rrbracket$.

 Maude provides several analysis methods,  including 
simulation by rewriting (command \lstinline[mathescape]{rew}), 
explicit-state
reachability analysis 
(\lstinline{search}),
 and model checking. 
Built-in  sorts
\code{Boolean}, \code{Integer}, and \code{Real} are defined for the corresponding 
 SMT theories. Rational constants of sort \code{Real} are written
\code{$n$/$m$} (\eg{} \code{0/1}).
Maude
uses two theory transformations 
to implement
symbolic rewriting as ``standard'' rewriting (\cite{rocha-rewsmtjlamp-2017}), thus opening the 
possibility of 
using standard Maude's commands 
on constrained terms. 
%

\section{Folding Narrowing with delayed SMT Constraints}\label{sec:narrowing}

Narrowing 
was originally defined as a method for equational unification by
the seminal work of \cite{fay1978first}
but emerged as a symbolic model checking method in \cite{ThatiMeseguer05}.
Narrowing is efficiently implemented in Maude together with unification modulo axioms and variant unification~\cite{DEEM+20}. This makes narrowing a very powerful verification technique used in many fields such as protocol analysis (see Maude-NPA, \cite{EMM09}), theorem proving (see NuITP, \cite{NuITPPPDP2024}) or deductive model checking (see \cite{BELMSJLAMP26}). 

We extend the folding narrowing
of 
\cite{lopezrueda2023-jlamp,DBLP:conf/ftscs/0001LS23}
(but we omit the irreducibility constraints  for simplicity) 
to \emph{delayed folding narrowing},
where SMT expressions are extended with new equational symbols
that become evaluable after proper instantiation. This generalization allows for  more 
expressive conditions in rewrite rules,
as the constraint \lstinline{noRecv(Prs)}   
stating  that no other agent in the set \texttt{Prs} is expecting a message: 
\lstinline|crl Prs + p(x) + clk(u) => Prs + p(x) + m(t) + clk(u) if t > u and noRecv(Prs)|.

\begin{definition}[SMT Extension]\label{def:smtextension}
For an equational theory $(\Sigma,E)$
with a \emph{built-in theory} $(\Sigma_0,E_0)$,
an \emph{SMT extension} $(\Sigma^\sharp_0,E^\sharp_0)$
is defined as 
(i)~$\Sigma_0 \subseteq \Sigma^\sharp_0 \subseteq \Sigma$,
(ii)~each sort from $\Sigma_0$ is minimal in $\Sigma^\sharp_0$,
(iii)~each symbol from $\Sigma^\sharp_0$ has no  subsort-overloaded typing in $\Sigma_0$,
(iv)~each symbol from $\Sigma^\sharp_0\setminus\Sigma_0$
is defined at the kind level, 
and
(v)~for $\Sigma_0$-terms $t_1$ and $t_2$,
$\mathcal{T}_{\mathcal{E}^\sharp_0} \models t_1 = t_2 \Leftrightarrow \mathcal{T}_{\mathcal{E}_0} \models t_1 = t_2$. 
\end{definition}

We extend the abstraction of built-ins in Section~\ref{sec:prelim}
so that any non-variable extended built-in subterms of a constraint $\phi$ is 
replaced by distinct built-in variables in $\phi^\sharp$
such that $\phi^\sharp$ is a pure SMT expression.

\begin{definition}[Extended SMT Expression Abstraction]
An abstraction of built-ins
of an extended SMT expression $\phi \in T_{\Sigma^\sharp_0}(X^\sharp_0)$
is a pair $(\phi^\sharp, \sigma^\sharp)$
of 
a constraint $\phi^\sharp \in T_{\Sigma_0}(X_0)$
and
a substitution $\sigma^\sharp : X_0 \to T_{\Sigma^\sharp_0}(X^\sharp_0)$
such that
$\phi = \phi^\sharp \sigma^\sharp$
and  $\phi^\sharp$  contains no duplicate variables in $X_0$.
The new equalities are
$\Psi^\sharp_{\sigma^\sharp} = \bigwedge_{x \in \mathit{dom}(\sigma^\sharp)} (x = x \sigma^\sharp$).
\end{definition}

For example, given an extended SMT expression
$\phi=``\code{T >= 0/1}\allowbreak\ \code{and}\allowbreak\ \code{T' >= 0/1}\allowbreak\ \code{and}\allowbreak\ \code{T' >= T}\allowbreak\ \code{and}\allowbreak\ \code{$\mathit{mte}$($t$, T')}$'',
where the constraint $\mathit{mte}(t,T')$ is the only non valid SMT expression and cannot be evaluated until it becomes completely instantiated,  
the abstraction $\phi^\sharp$
removes this subexpression by a new variable: 
$\phi^\sharp=``\code{T >= 0/1}\allowbreak\ \code{and}\allowbreak\ \code{T' >= 0/1}\allowbreak\ \code{and}\allowbreak\ \code{T' >= T}\allowbreak\ \code{and}\allowbreak\ \code{W}$''.

\begin{lemma}%
Let $\phi \parallel t$ be a constrained term,  
$(\phi^\sharp, \sigma^\sharp)$ be an
abstraction of built-ins for the extended SMT expression $\phi$
and assume that $\mathit{dom}(\sigma^\sharp) \cap \ovars{\phi \parallel t} = \emptyset$. Then: 
 (1) 
$\llbracket \phi \parallel t \rrbracket
= 
\llbracket \phi^\sharp \wedge \Psi_{\sigma^\sharp} \parallel t \rrbracket$; and
 (2) If 
the implication $\phi \Rightarrow \phi^\sharp$ holds,
then
$
\llbracket \phi \parallel t \rrbracket
\subseteq
\llbracket \phi^\sharp \parallel t \rrbracket
$.
\end{lemma}


Below, the one-step narrowing relation
of 
\cite{Lopez-RuedaEM22-Canonical}, \cite{lopezrueda2023-jlamp} and \cite{DBLP:conf/ftscs/0001LS23}
is extended to handle extended SMT expressions.

\begin{definition}[Delayed SMT Canonical  Narrowing]\label{def:canonical-narrowing}
    Let $(\Sigma,E,L,R)$ be a topmost order-sorted rewrite theory  including a 
    built-in theory $(\Sigma_0,E_0)$,  and an SMT extension $(\Sigma^\sharp_0,E^\sharp_0)$. The \emph{narrowing relation with SMT constraints and delayed expressions}  holds between $\phi~\parallel~t$ and $\phi'~\parallel~t'$, denoted  
$\phi~\parallel~t~\leadsto_{\alpha,R,E}~\phi'~\parallel~t'$ iff there exists $l : q\to r\mbox{ \textbf{if }}\psi\in R$, which we always assume renamed, so that $\ovars{\phi~\parallel~t}\cap(\ovars{r}\cup\ovars{q}\cup\ovars{\psi})=\emptyset$, and a unifier $\alpha\in\mathit{CSU}(t = q)^{W}_{E}$, where $W=\ovars{\phi~\parallel~t}\cup\ovars{r}\cup\ovars{q}\cup\ovars{\psi}$, 
       $(\phi' \parallel t') = (\phi\alpha \wedge \psi\alpha \parallel r\alpha)$,
       and
       $\phi'^\sharp$ is satisfiable.
\end{definition}


In \cite{DBLP:conf/ftscs/0001LS23}, we provided 
a subsumption relation on patterns $\phi \parallel t$ that we adapt here to extended SMT expressions.
Note that this subsumption relation  was already presented in \cite{DBLP:journals/jlap/Meseguer20} and used in \cite{ftscs-journal} to produce a finite (when possible) state graph for rewriting with SMT . 

\begin{definition}[Extended SMT Expression Subsumption]
Let $U_1=\phi_1 \parallel t_1$ and $U_2=\phi_2 \parallel t_2$. 
We write $U_2 \sqsupseteq^\sharp_{E} U_1$, meaning that $U_2$ is more general than $U_1$, 
if there is a substitution $\theta$ 
such that 
$t_1 =_{E} t_2\theta$
and the implication $\phi^\sharp_1 \Rightarrow (\phi_2\theta)^\sharp$ holds.  
\end{definition}

A state graph 
$\mathit{Post}^{\ast}_{{\cal K}\leq}(I)$
is defined in
\cite{DBLP:journals/jlap/Meseguer20}, \cite{lopezrueda2023-jlamp} and \cite{DBLP:conf/ftscs/0001LS23}
being parametric on a subsumption relation $\leq$
that we now instantiate to $\sqsupseteq^\circ_{E}$.

\begin{theorem}[Delayed Folding Narrowing]
    Let $\mathcal{R}= (\Sigma,E,L,R)$  be a topmost order-sorted rewrite theory including  a 
    built-in theory $(\Sigma_0,E_0)$ and an SMT extension
    $(\Sigma^\sharp_0,E^\sharp_0)$. Let 
${\cal N}({\cal R}) = ({\cal P},\leadsto_{R,E})$
be the state graph
such that ${\cal P} = T_{\Sigma}(X) \times T_{\Sigma^\sharp_0}(X_0)$
with 
a set of initial states $I \subseteq {\cal P}$. Then, 
the preorder $\sqsupseteq^\sharp_{E}$ 
is a folding preorder
for the state graph $\mathit{Post}^{\ast}_{{\cal N}({\cal R})\sqsupseteq^\sharp_{E}}(I)$.
\end{theorem}

If a \emph{finite} folding state graph does \emph{not} satisfy an invariant,
then there exists an \emph{error} state $s \in \mathit{Post}^{\ast}_{{\cal N}({\cal R})\sqsupseteq^\sharp_{E}}(I)$ that violates the invariant.
Because the error state $s$ is again reachable from $I$ in the original state graph,
a \emph{concrete counterexample}  can be constructed by repetitively traversing the cycles in the folded state graph.

In practice, checking whether  $\phi_1 \parallel t_1 \sqsupseteq^\sharp_{E} \phi_2 \parallel t_2$ 
amounts to determining if (i) $t_2$ is an instance of $t_1$, \ie{}
$\exists \theta: t_1\theta =_{E} t_2$, and then if 
(ii) $\neg(\phi_2^\sharp \Rightarrow (\phi_1\theta)^\sharp)$ is unsatisfiable (and hence, 
$\llbracket \phi_1 \parallel t_1 \rrbracket \supseteq \llbracket \phi_2 \parallel t_2 \rrbracket$). 
Following \cite{ftscs-journal}, we check the unsatisfiability 
of $\neg(\exists \vec{x}. \phi_2^\sharp \Rightarrow (\phi_1\theta)^\sharp)$, where
$\vec{x}$ is the set of SMT variables 
occurring in $(\phi_1\theta)^\sharp$ and $\phi_2^\sharp$ but not in  $t_1\theta$ and $t_2$. The reason is that  
the information on those variables, not occurring in the term,  is ``irrelevant'' for checking the subsumption
relation. 


In ``standard'' Maude, only unconditional rules for narrowing are possible and they are declared with 
the label ``narrowing'' as in 
\lstinline[mathescape]{rl $t$ => $t'$ [narrowing]}. 
Maude  supports folding narrowing-based reachability analysis 
via the command
%
\lstinline[mathescape]{fold vu-narrow [$n$,$d$] $t_1$ =>* $t_2$}
%
\noindent where $n$ denotes the number of solutions and $d$  the maximum depth. The delayed folding narrowing presented here  is not part of Maude
but an implementation in Maude using meta-level features, and it is available in the companion repository.
The conditional SMT rules used for the delayed folding narrowing 
are declared as follows:
\lstinline[mathescape]{crl $t$ => $t'$ if $\phi$ [nonexec]}
where $\phi$ is an extended SMT expression
(internally, the keyword \lstinline{narrowing} is added).
We implemented a new command 
\lstinline[mathescape]{fold vu-narrow [$n$,$d$] $t_1$ =>* $t_2$ such that $\phi$}
where 
$\phi$ is an extended SMT expression that we illustrate in the next section. 
We note that 
this command extends the one in Maude  with the ``\lstinline{such that}'' part (for extended and ``standard''
conditions).

\section{Logical Real-Time Rewrite Theories}\label{sec:theory} 

This section shows how real-time systems can be analyzed using the
narrowing-based framework introduced in the previous section. We
show how to specify the system behavior as a \emph{logical real-time rewrite
theory}~$\rttheory$. When initial states are specified as ground terms, such a
theory can be executed and analyzed in Maude with SMT. More interestingly,
$\rttheory$ enables narrowing-based analyses that combine SMT constraints with
 \emph{delayed} constraints. By allowing logical variables in the
initial state, $\rttheory$ supports richer analyses of real-time systems
as evidenced by the case studies shown in Sections~\ref{sec:fischer} and ~\ref{sec:other-app}.

\paragraph{Logical real-time rewrite theories.}

Real-time systems can be naturally modeled in rewriting logic as
\emph{real-time rewrite theories} (\cite{OlvMesTCS}), which are
parametric in the (discrete or dense) time domain. The idea is
that ordinary rewrite rules model \emph{instantaneous}
transitions (changes in the state of the system), and \emph{tick} rewrite rules model 
the passage of time. 
We define a \emph{dense} time domain using Maude's built-in sorts
\texttt{Boolean} and \texttt{Real} for the corresponding SMT theories:

\begin{maude}
fmod TIME-DOMAIN is protecting REAL .
  sort Time .  subsort Real < Time .     
\end{maude}

The system (\lstinline{sort System} below) is built from a multiset (\lstinline{sort MSTOs}) of \emph{timed objects} (TOs) (\lstinline{sort TimedObject}) together with
the global clock of the system:

\begin{maude}
fmod LRT-THEORY is protecting TIME-DOMAIN .
  sorts TimedObject MSTOs System .   subsort TimedObject < MSTOs .
  op empty : -> MSTOs [ctor] .
  op __ : MSTOs MSTOs -> MSTOs [ctor assoc comm id: empty] .
  op {_} in time_ : MSTOs Real -> System [ctor] .
 \end{maude}
The empty syntax (\lstinline|op __|, which is an associative, 
commutative and with identity operator) represents multiset union.
Due to the subsort relation \lstinline|TimedObject < MSTOs|, 
a timed object $T_o$ is also a singleton (a term of sort \lstinline|MSTOs|). 
A system is a term of the form 
\lstinline[mathescape]|{ $M$ } in time $T$| where $M$ is a multiset of TOs
and the value of the global clock is $T$. 


\begin{example}[Timed Objects]\label{ex:tos}
    Parametric timed automata (PTA, see e.g., \cite{Andre21}) extends timed
    automata with system parameters whose values are initially unknown:
\begin{maude}
fmod PTA is including LRT-THEORY .
  sort Location .      ops l1 l2 ... ln :   -> Location    [ctor] .
  op <loc:_,x:_,y:_>   : Location Real Real -> TimedObject [ctor] .
  op <gamma:_,delta:_> : Real Real          -> TimedObject [ctor] .
\end{maude}
The TO  \lstinline[mathescape]|<loc: $l_i$, x: $r_x$, y: $r_y$>|
specifies that the automaton is currently at location $l_i\in\{l_1,\cdots,l_n\}$, where its two clocks have values $r_x$ and $r_y$, respectively. 
As explained below, different from other approaches as~\cite{ftscs-journal},
$r_x$ and $r_y$ are indeed \emph{timestamps} and the actual value of the clock $x$
is given by the expression $T-r_x$ where $T$ is the global clock (similarly for clock $y$). 
The TO \lstinline[mathescape]|<gamma: $\gamma$, delta: $\delta$>|
represents the two parameters of this PTA (both expressions $\gamma$ and $\delta$ of sort \lstinline{Real}). 

%
%
\end{example}

\noindent Rewrite rules are topmost and we distinguish
two kinds of rules: \emph{instantaneous} and \emph{tick} rules. The former represents a change
in the state of the system. For instance, the rule 

\begin{maude}
vars X Y T : Real .  var SYS : MSTOs .
crl [move] : { <loc: $\textcolor{red}{l_1}$, x: $\highlight{X}$, y: Y > SYS } in time T =>
             { <loc: $\textcolor{red}{l_2}$, x: $\highlight{T}$, y: Y > SYS } in time T if $\phi$(X,Y,T) [nonexec] .
\end{maude}

\noindent states that whenever the current location is $l_1$
and  $\phi$ holds, the automaton moves to  
 $l_2$ and \emph{resets} the clock $x$, by taking as a timestamp the current value of the global clock.



Tick rewrite rules model a time elapse, advancing the global clock by
an (undetermined) amount \lstinline{T' - T} (where \lstinline{T'} does not occur in the LHS):
\begin{maude}
crl [tick] : {$t$} in time T => {$t$} in time T' if T' >= T and $\mathit{mte}$($t$, T') [nonexec] .
\end{maude}
%

The operator \lstinline{op mte : MSTOs Time -> [Boolean]}
 determines an upper bound for  $T'$  and it is an extension of the SMT domain (see Def.~\ref{def:smtextension}). This constraint is delayed and becomes evaluable only when properly instantiated as in CLP. 
The symbol $\mathit{mte}$  is only declared in theory \lstinline{LRT-THEORY} 
and needs to be defined 
equationally by a user module. Since $\mathit{mte}$ 
will be evaluated by rewriting when properly instantiated, it can be specified using regular equations in Maude without the keyword \lstinline{variant}.

\begin{definition}[Logical Real-time Rewrite Theory]\label{def:ltheory}
We call $\rttheory$ a logical real-time rewrite theory when it extends the theory 
\lstinline{LRT-THEORY} with equations defining the function $\mathit{mte}$, and 
topmost rewrite rules of the form
(assuming that \lstinline{SYS} is a variable of sort \lstinline{MSTOs}, \lstinline{T},\lstinline{T'} are 
        of sort \lstinline{Real}): 
        \begin{maude}
crl [$\ell$] : {$to_1$ $\cdots$ $to_n$ SYS} in time T =>
          {$to'_1$ $\cdots$ $to'_j$ $to_{n+1} \cdots to_m$ SYS} in time T if $\phi$  [nonexec] .
crl [tick] : { SYS } in time T => { SYS } in time T' if T' >= T and mte(SYS, T') [nonexec] .
\end{maude} 
In the first rule,  $j\leq n$, 
	 all the variables on the RHS appear in the LHS,  and
all SMT positions in the LHS are distinct variables (of sort \lstinline{Real}). 
The TOs $to'_1\cdots to'_j$
are updated versions of the TOs
$to_1\cdots to_j$,
the TOs $to_{n+1}\cdots to_m$ are newly created TOs, 
and the TOs $to_{j+1}\cdots to_{n}$ were removed. 
\end{definition}

\begin{figure}[t]
    \centering
    \begin{subfigure}[t]{0.48\textwidth}
        \centering
        \includegraphics[width=\linewidth]{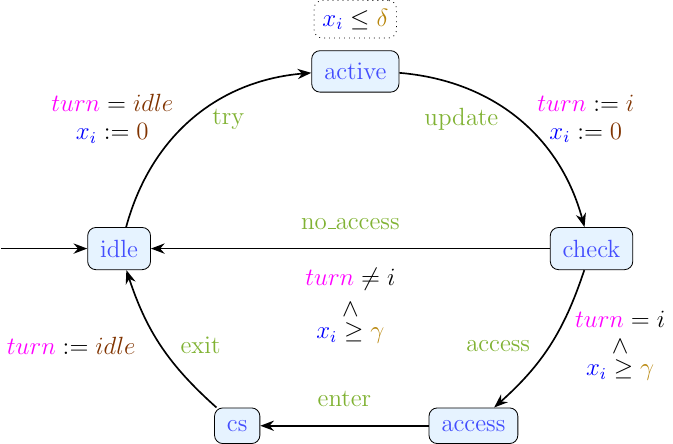}
        \caption{}
        \label{fig:fischer}
    \end{subfigure}%
    \hfill
    \begin{subfigure}[t]{0.48\textwidth}
        \centering
        \includegraphics[width=\linewidth]{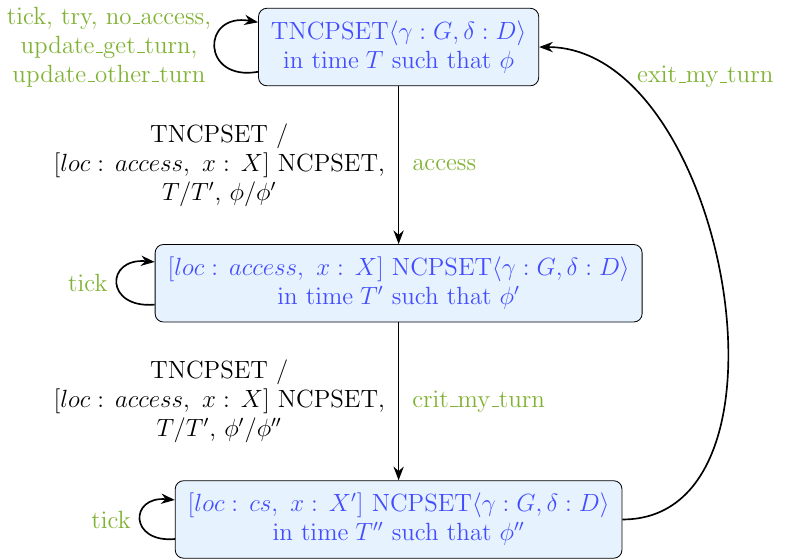}
        \caption{}
        \label{fig:fischer-folding}
    \end{subfigure}

    \caption{(a) Fischer protocol. Automaton for process $P_i$, with identifier $i$ and local clock $x_i$. 
    \texttt{turn} is a global variable shared by all processes, and $\delta$ and $\gamma$ parameters. (b) Search space for the Fischer protocol when using folding narrowing with delayed SMT constraints.}
    \label{fig:fischer-both}
\end{figure}

\subsection{Fischer Protocol}\label{sec:fischer}
This subsection 
presents a logical real-time rewrite theory for a timed mutual
exclusion protocol, a benchmark for parametric timed automata (\cite{Andre21}).
This system has been verified in~\cite{ftscs-journal} using rewriting modulo
SMT when the initial state, a ground term, includes only two processes. In the
following we show how to verify it when considering an arbitrary number of
processes and arbitrary parameters $\gamma$
and $\delta$. 

  Figure~\ref{fig:fischer} shows 
  the PTA corresponding to one process executing the 
   protocol. Each process 
$i$ has its own clock \styleclock{x_i}. There are two
parameters, $\delta$ and $\gamma$, which represent the upper and lower time bounds of the protocol. The global variable \styledisc{turn}, shared by all the processes,  
records which process is to enter its critical section.

A process is initially at location \styleloc{idle} and  can try to enter the
critical section with the transition \styleact{try}, restarting its clock. This
is possible only if \styledisc{turn} is equal to \styleconst{idle}. The process
remains at location \styleloc{active} for at most $\delta$ time units
(this is the invariant at location \styleloc{active}) and then can perform an
\styleact{update}, again resetting  its clock, and recording its number in
variable \styledisc{turn}. When the minimum time $\gamma$ has
elapsed, the process can test whether it is still its turn. If this is the
case, the process gets \styleact{access} to the critical section. Otherwise, it
has to return to location  \styleloc{idle}. When  access is granted, the
process can  \styleact{enter} the  critical section to later \styleact{exit}.
On doing that, the process sets \styledisc{turn} to \styleconst{idle}. 

The proposed specification  distinguishes between processes that hold the turn and those that do not. Additionally, a distinction is also made between processes that are located in critical locations (\styleloc{access} and \styleloc{cs}) and those that are not (\styleloc{idle}, \styleloc{active}, \styleloc{check}). To this end, a hierarchy of sorts is defined, whose common supersort is precisely \lstinline{TimedObject}, so that all processes in the system are TOs: 

\begin{maude}
sorts NCProc CProc Proc TNCProc TCProc TProc .
subsorts NCProc CProc < Proc < TimedObject .
subsorts TNCProc TCProc < TProc < TimedObject .
op <`loc:_,x:_> : NCLocation Real -> NCProc  [ctor] .
op <`loc:_,x:_> : CLocation Real  -> CProc   [ctor] .
op [loc:_,x:_] : NCLocation Real  -> TNCProc [ctor] .
op [loc:_,x:_] : CLocation Real   -> TCProc  [ctor] .
\end{maude}

The sort \lstinline{NCProc} defines a process without the turn 
that it is located in non-critical locations (\styleloc{idle}, \styleloc{active}, \styleloc{check}), whereas the sort \lstinline{CProc} defines a process also without the turn but located in critical locations (\styleloc{access}, \styleloc{cs}). Both are subsumed by a supersort \lstinline{Proc}, which therefore unifies processes without the turn regardless of whether they are in critical or non-critical locations.
Likewise, a hierarchy is defined for processes that do hold the turn, 
using the sorts \lstinline{TNCProc} (resp. \lstinline{TCProc})
for processes with the turn in non-critical (resp. critical)  locations.
The sort that unifies both in this case is \lstinline{TProc}. 
Syntactically, this distinction is reflected by the object constructors: processes written as \lstinline{<...>} represent processes that do not hold the turn, whereas processes written as \lstinline{[...]} denote processes that currently hold the turn. In this way, the presence or absence of the turn is directly encoded in the object notation.

With these sorts, the specification  becomes very natural, generally creating one rule per system transition, although in some cases it is necessary to split it into two rules to capture certain distinctions. For instance, when a process is in location \styleloc{check} and reaches or exceeds time $\gamma$ (see Figure~\ref{fig:fischer}), two different situations may arise depending on whether or not the process holds the turn  at that moment:

\begin{maude}
crl [no-access] : 
    { < loc: $\highlight{check}$, x: X > < gamma: GAMMA, delta: DELTA > TPSET } in time T => 
    { < loc: $\highlight{idle}$,  x: X > < gamma: GAMMA, delta: DELTA > TPSET } in time T
if (T - X >= GAMMA) = true .
crl [access] : 
    { [ loc: $\highlight{check}$, x: X ]  < gamma: GAMMA, delta: DELTA > PSET } in time T =>  
    { [ loc: $\highlight{access}$, x: X ] < gamma: GAMMA, delta: DELTA > PSET } in time T
if (T - X >= GAMMA) = true .
\end{maude}

In  \lstinline{no-access}, the process to which the location transition is applied attempts to move to \styleloc{access}, but, since it does not hold the turn, it is forced to move to \styleloc{idle}. In contrast, in rule \lstinline{access}, the process does hold the turn and is therefore allowed to move to \styleloc{access}. Depending on the rule, the multiset variables \lstinline{TPSET} (sort \lstinline{TProcSet}, for multisets of \lstinline{TProc}s) and \lstinline{PSET} (sort \lstinline{ProcSet}, for multiset of \lstinline{Proc}s) are used.

The advance of time in the system is captured by a single rule as the one in~\Cref{def:ltheory},
where 
the function $\mathit{mte}$  operates over the different objects of the system. For instance,  
it is defined to enforce the invariant at location $\styleloc{active}$ (other cases are similar): 



%

\begin{maude}
op mte : Location Real Real Real Real -> [Boolean] .
eq mte(active, X , T, DELTA, GAMMA) = (T - X <= DELTA) .
\end{maude}

Now we can use our folding narrowing with delayed SMT constraints to verify properties of the system. 
If the initial state is a ground term containing two  \styleloc{idle} processes,
the safety property is violated. 

\begin{maude}
{fold} vu-narrow [1] in FISCHER : { < loc: idle, x: X > < loc: idle, x: X >... } 
=>* { < loc: cs, x: X' > [ loc: cs, x: Y' ] SYS } in time T' 
such that T >= 0 and DELTA >= 0 and GAMMA >= 0 .

Solution 1, state 121:
...
\end{maude}

The returned accumulated constraint shows that the state  was reached assuming that $\delta \geq \gamma$. 
If we launch the same command with initial constraint $\gamma > \delta$, 
we obtain \lstinline{No solution}. This means
 that the search space becomes finite (due to folding), and narrowing is able to check
 that none of the reachable states violate the property when $\gamma > \delta$. 
 


%


The symbolic capabilities of our narrowing also allow for more general queries than rewriting, using free variables in the initial state to represent an arbitrary number of processes. In any reasonable initial state, we expect that those processes are in non-critical locations and,
at most one of those, have the turn. These requirements for the initial logical state can be naturally captured by the defined  hierarchy of sorts:


%
%

\begin{maude}
{fold} vu-narrow [1] in FISCHER :
{ TNCPSET < gamma: GAMMA, delta: DELTA > } in time T =>* ...
such that T >= 0 and DELTA >= 0 and GAMMA >= 0 and GAMMA > DELTA  .

No solution.
\end{maude}

The result of this command establishes the mutual exclusion property for an arbitrary number of processes 
that are initially  in non-critical sections. Due to the careful design of sorts  the search space contains only 3 states, as shown in Figure~\ref{fig:fischer-folding}.
The initial symbolic state, due to the use of the logical variable \lstinline{TNCPSET}, 
subsumes an infinite set of states: all those with an arbitrary number of processes in any non-critical location, including those where at most one process holds the turn. From Figure~\ref{fig:fischer-folding}, multiple states are generated using the rules \styleact{tick}, \styleact{try}, \styleact{no\_access}, \styleact{update\_get\_turn}, and \styleact{update\_other\_turn}. However, all of these states are subsumed by the initial state itself. Only the rule \styleact{access} produces a new state that cannot be subsumed as it contains a critical process in location \styleloc{access}. From this new state, the previous rules cannot be applied, except for \styleact{tick}. Some of these rules, such as \styleact{try}, cannot be applied because a process already holds the turn. Others are prevented by the constraint $\gamma > \delta$. The rule \styleact{enter}, however, can be applied, producing another new state that contains a critical process in location \styleloc{cs}. From this last state, only \styleact{tick} or \styleact{exit\_my\_turn} can be applied, returning to the initial state.

\subsection{Reachability and Synthesis}\label{sec:other-app}

The previous section explored the use of folding narrowing with delayed SMT expressions to prove a system invariant by checking  the unreachability of certain states. This section explores the use of variables in the initial state to \emph{synthesize} part of the system so that a reachability property holds. We illustrate this idea with the classical dining philosophers problem (TDP), extended with clocks and timed parameters, together with a lackey coordinating philosophers' access to the dining room~\citep{DBLP:conf/ppdp/0001OPPS24}.

\begin{figure}
\begin{tblr}{
        colspec = {Q[c,m] Q[c,m]}, 
    }
        {
            \includegraphics[width=0.3\textwidth]{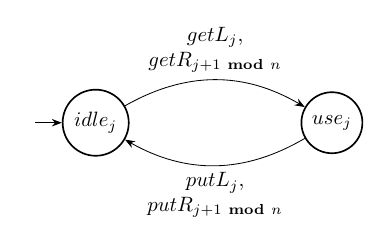} \\
            \includegraphics[width=0.4\textwidth]{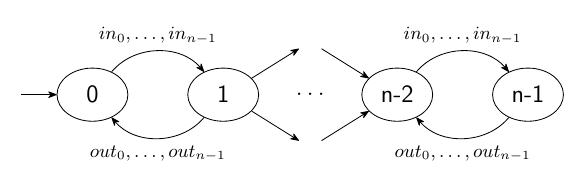}
        } 
        & 
        \includegraphics[width=0.45\textwidth]{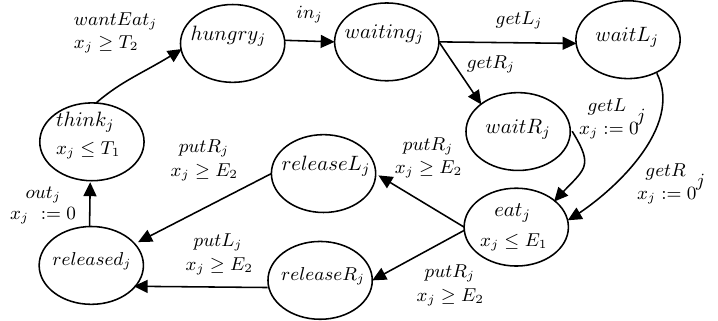} 
    \end{tblr}
\vspace{-.5cm}
\caption{Automata for the $Fork_i$, the lackey and $\mathit{Philosopher}_i$. \label{fig:tdp}}
\vspace{-.5cm}
\end{figure}

The automata for the TDP are given in~\Cref{fig:tdp}.
Each philosopher $j$ uses a clock $x_j$, he has to think at least $T_2$ and at most $T_1$
time-units, and he is supposed to eat for at most $E_1$ and at least $E_2$ time-units (see the invariants and guards in the figure). 
$T_1,T_2,E_1$ and $E_2$ are parameters of the system. The philosopher
$j$ is allowed to move to the state $\mathit{waiting}_j$ only if the 
lackey offers the corresponding synchronized action $in_j$. Similarly, 
the lackey controls the exit of the philosophers by using the synchronized action $out_j$.
Before eating, the philosophers have to grab the needed forks (synchronized actions   $getL_j$, $getR_j$), and release the forks after eating
(synchronized actions $putL_j$ and $putR_j$). 

One interesting synthesis problem for the TDP is to
find a controller, i.e., an automaton for the lackey,
such that some of the philosophers have access to  the dining room before a given deadline.
The automaton for the lackey in~\Cref{fig:tdp} is one
of the possible solutions for this synthesis problem. 

The model of this TDP 
as a logical real-time rewrite theory follows the same principles described in the previous sections.
We therefore refer to the companion repository for the complete specification,
and below we highlight some of the key design choices.

We consider an \emph{interpreter},
where the specification of the PTAs is given as a term $net$ of sort \lstinline{Network},
which in turn is a set of terms of sort \lstinline{Automata}. Hence, we   consider only two rewrite rules: \lstinline{tick}, as before, and a rule that, given the current state and $net$, rewrites to a new state by resetting clocks and moving to the next locations accordingly.
As mentioned above, the controller/lackey uses actions that can synchronize with the transitions of the philosophers.
In particular, actions $in(j)$ and $out(j)$ are used by the lackey to allow philosopher $j$ to enter and exit the dining room. 
We therefore distinguish between \emph{basic} actions, which are under the control of the automaton being synthesized, and the remaining actions (e.g., those used by philosophers and forks to synchronize):

\begin{maude}
sorts Action BAction .  subsort BAction < Action .
ops in out : Nat -> BAction .  ops getL getR putL putR : Nat -> Action  .  
\end{maude}

Transitions in PTAs may have guards and reset operations, and locations
may have invariants. A full synthesis problem may consider an arbitrary automaton for the lackey
with those features. However, as shown 
in, e.g., \citep{DBLP:conf/ppdp/0001OPPS24}, the state space of the TDPs
grows very quickly, and it is out of the question to consider this full generality. Hence,
we consider ``standard'' automata, and
\emph{basic} automata (those to be synthesized) whose  transitions do not have invariants or guards:

\begin{maude}
sorts StateDef BStateDef Automaton BAutomaton  .  subsort BAutomaton < Automaton .
op sync_goto_ :  BAction Location -> STransition [ctor] . --- No guards/resets
op @_inv_:_ : Location Constraint SetTransition -> StateDef [ctor] .
op @_:_ : Location SetSTransition -> BStateDef [ctor] . --- Def without invariant
op <_|_> : AutoId SetStateDef  -> Automaton    [ctor] . --- Standard automaton
op <_|_> : AutoId SetBStateDef -> BAutomaton   [ctor] . --- Basic automaton 
\end{maude}

Consider the execution of the following command:
\begin{maude}
vu-narrow [1] in TDP : {(< lackey | $\highlight{DEFS:SetBStateDef}$ >, Forks(4), Phils(4))} ... 
=>* ... locs: (LOCS:SetLocs,  phil(0) @ waiting(0), phil(2) @ waiting(2)) ... 

Solution 1, state 1714:
accumulated substitution: DEFS:SetBStateDef --> ($\$$6958:SetBStateDef ,
    @ $\$$6959:Location : $\$$6960:SetSTransition, sync in(s(2)) goto $\$$6961:Location ),
    @ l(0): $\$$6962:SetSTransition ,sync in (0) goto $\$$6959:Location...
\end{maude}

\noindent
The automaton for the lackey is initially undefined (variable \lstinline{DEFS}, a set of ``basic'' state definitions), while the automata for the $4$ philosophers and forks are given. The pattern of the query specifies that we are looking for configurations in which philosophers~$0$ and~$2$ have already entered the dining room.
The reported answer constrains \lstinline{DEFS} to contain at least three locations (\lstinline{l(0)}--the initial location--, \lstinline[mathescape]{$\$6959$} and \lstinline[mathescape]{$\$6961$}, where the numbers are generated during narrowing to guarantee freshness), possibly along with other ``basic'' definitions (variable \lstinline[mathescape]{$\$6958$}). 
Moreover, there is a transition from \lstinline{l(0)} that offers the action \lstinline{in(0)} and moves to location \lstinline[mathescape]{$\$6959$}, and location \lstinline[mathescape]{$\$6959$} offers the action \lstinline{in(2)}. 
This therefore constitutes one of the possible synthesized automaton for the lackey that guarantees the reachability property.

Once the automaton of the lackey has been synthesized, a
 further \lstinline|vu-narrow| command
can be used to check that the resulting system satisfies the expected safety  property
as we did in the previous section:
there is no reachable configuration violating the invariant, i.e., there is no configuration 
where two consecutive 
philosophers are both eating.


\section{Related Work}\label{sec:related}

Maude provides different narrowing-based reachability features (\cite{DEEM+20,maude-manual}), and 
several extensions have been proposed.
In \cite{DBLP:journals/jlap/Meseguer20},
a theoretical development of narrowing with constraints is provided.
We also worked on conditional narrowing algorithms, including narrowing with SMT constraints \citep{Lopez-RuedaE22-SMT} and narrowing for variant-based conditional rewrite theories \citep{DBLP:conf/icfem/Lopez-RuedaE22}, and the combination of both {\rm\citep{lopezrueda2023-jlamp}}.
In \cite{DBLP:conf/ftscs/0001LS23}, we also provided a canonical narrowing with SMT constraints and irreducibility constraints that is able to obtain a finite search space. 
Most of these works include 
the folding narrowing idea, 
based on the unfold/fold program manipulation and transformation approach of 
\cite{Burstall-Darlington-77}.
The new delayed folding narrowing relation extends all these previous works. Furthermore, it has been extended to irreducibility constraints but it is outside the scope of present paper.
The new delayed folding narrowing 
complements 
\cite{BELMSJLAMP26}.

Due to the extra variable $T'$ in the RHS, tick rules are 
non-executable and the analysis of real-time systems in
rewriting logic has traditionally relied on \emph{explicit-state}
executions, where tick rules are executed using specific
\emph{time sampling strategies} to instantiate  $T'$. This is 
in general neither sound nor complete in dense time settings (see~\cite{DBLP:journals/jlap/OlarteO25}). 
Recently, we have shown that using rewriting modulo
SMT (\cite{rocha-rewsmtjlamp-2017}), it is possible to perform sound and complete
analyses  for PTAs, 
time Petri nets,
and general real-time rewrite theories
(\cite{ftscs-journal,DBLP:journals/fuin/AriasBOOP24,Bae2026}
 --see also these references for other tools and methods
for the analysis of real-time systems.  
Intuitively,
instead of ``guessing'' the value of $T'$, the constraints required for applying
the tick rule are \emph{accumulated} and then checked for satisfiability. 

In previous works, the initial query always contains only SMT variables (e.g., PTA parameters), while the number of automata or processes is fixed. This paper is the first attempt to analyze such systems with an arbitrary number of processes or agents. In particular, time-object sorts were carefully designed to ensure termination, and timestamps model clock resets. This allows tick rules to advance only the global clock, without modifying the system state after the transition, unlike the $\mathit{timeEffect}$ function in Real-Time Maude, which cannot be used here because the system state may contain logical variables.

\section{Concluding Remarks}\label{sec:conc}

We proposed a new narrowing with delayed constraints that enables the verification of real-time systems in Maude when the initial state may include logical variables. This approach allows for verification tasks beyond those supported by state-of-the-art tools for PTAs and ``standard'' Maude. In the companion repository~\citep{tool}, we  include additional examples as other PTAs, and the specification of time Petri nets with SMT parameters in the timed transitions and logical variables in the initial marking. For those nets, we are able to synthesize the initial marking that ensures certain properties. 

Future work includes: the implementation of the proposed narrowing at the C++ level of Maude, thus improving efficiency; implementing disjunctive patterns for the initial state, facilitating reasoning about folding and reducing the design complexity of the sort hierarchy; combining delayed folding narrowing with variant or reachability conditions; and automating the synthesis of constraints by iteratively finding counterexamples 
and checking whether the ``negation'' of the output ensures the desired property. 

\textbf{Acknowledgments.} We would like to thank the anonymous reviewers for their very insightful comments on
an earlier version of this paper. The authors acknowledge support from the NATO Science for Peace and
Security Program through grant number G6133.
S.~Escobar and R.~L\'opez 
have also been supported 
by the grant PID2024-162030OB-100 funded by MCIN/AEI/10.13039/501100011033 and ERDF A way of making Europe,
and
by the grant CIPROM/2022/6 funded by Generalitat Valenciana.

\bibliographystyle{eptcs}
\bibliography{biblio}

\end{document}